\documentclass[twocolumn,letterpaper]{igs}

\usepackage{verbatim,xspace,amsmath,amssymb,bm,multirow,dashbox}
\usepackage{tikz}
\usetikzlibrary{arrows}
\usepackage{igsnatbib,lineno}

\usepackage[kw]{pseudo}
\pseudoset{left-margin=0mm,topsep=5mm,idfont=\texttt}

\usepackage{hyperref}
\hypersetup{pdfauthor={Ed Bueler},
            pdfcreator={pdflatex},
            colorlinks=true,
            citecolor=purple,
            linkcolor=red,
            urlcolor=blue,
            }

\newcommand\bu{\mathbf{u}}

\newcommand\bx{\mathbf{x}}

\newcommand{\grad}{\nabla}

\newcommand{\Divx}{\nabla_\bx \cdot}
\newcommand{\gradx}{\nabla_\bx}

\begin{document}

\title[Performance analysis of high-resolution ice sheet simulations]{Performance analysis of high-resolution \\ ice sheet simulations}

\abstract{Numerical ice sheet models compute evolving ice geometry and velocity fields using various stress-balance approximations and boundary conditions.  At high spatial resolution, with horizontal mesh/grid resolutions of a few kilometers or smaller, these models usually require time steps shorter than climate-coupling time scales because they update ice thickness after each velocity solution.  High-resolution performance is degraded by the stability restrictions of such explicit time-stepping.  This short note, which considers the shallow ice approximation and Stokes models as stress-balance end members, attempts to clarify numerical model performance by quantifying simulation cost per model year in terms of mesh resolution and the number of degrees of freedom.  The performance of current-generation explicit time-stepping models is assessed, and then compared to the prospective performance of implicit schemes.  The main result, Table \ref{tab:performancemodel}, highlights the key roles played by the algorithmic scaling of stress-balance and implicit-step solvers.}

\author{Ed Bueler}

\affiliation{Dept.~Mathematics and Statistics, University of Alaska Fairbanks, USA \\
E-mail: \emph{\texttt{elbueler\@@alaska.edu}}}

%\keywords{glacier flow,glacier modelling,ice-sheet modelling}
% suggested reviewers:  Dan Goldberg, William Lipscomb, Toby Isaac, Josefin Ahlkrona, Alex Jarosch, Doug Brinkerhoff

\maketitle

\sectionsize

\section{Introduction}

Numerical ice sheet (glacier) models with evolving ice geometry are now routinely used for addressing scientific questions such as quantification of future sea level rise from changes in the Antarctic \citep{Seroussietal2020} and Greenland \citep{Goelzeretal2020} ice sheets, interpretation of the paleoglacial record \citep{Weberetal2021}, and evaluation of long-term glacial erosion rates \citep{SeguinotDelaney2021}, among other applications.  It is generally accepted that horizontal mesh (grid) cells must be smaller than about 10 km in order to generate valid results, but narrow outlet glacier flows need yet finer resolution.  Whether using local mesh refinement \citep[for example]{Fischleretal2022,Hoffmanetal2018} or not, resolutions of two or one kilometers \citep{SeguinotDelaney2021} or less \citep{Aschwandenetal2019} are increasingly used for science at ice sheet scale.

Current-generation ice sheet models use a variety of stress balances, from the simplest shallow ice approximation (SIA), through ``hybrid'' \citep{Robinsonetal2022,Winkelmannetal2011} and higher-order balances, up to the non-shallow and non-hydrostatic Stokes approximation.  With very few exceptions, however, current time-stepping models alternate between solving the stress balance for velocity, using the geometry determined by the previous time step, and then updating the geometry using the just-computed velocity field.  Therefore ice thickness and surface elevation, functionally-equivalent variables for this purpose, are updated after velocity is fixed.  Note that the interaction between the ice sheet and the surrounding climate occurs during this geometry-update operation, via the mass continuity or surface kinematical equations \citep{GreveBlatter2009}.  Such climatic coupling occurs through surface mass balance, sub-shelf (basal) mass balance, and calving processes in particular.

These current-generation schemes implement \emph{explicit} time-stepping for the coupled mass and momentum system which describes the dynamical evolution of ice sheets.\footnote{Confusingly, various ``semi-implicit'' and even ``fully implicit'' designators appear in the literature for explicit schemes which use velocity computed from the geometry at the start of the time step \citep[for example]{Chengetal2017}.}  At least for simpler, textbook partial differential equation problems, the conditional stability of such explicit time-stepping schemes is well-understood \citep{LeVeque2007}.  Stability conditions of explicit SIA models have also been understood for some time \citep{HindmarshPayne1996}, but recent studies have focussed on the stability conditions of explicit hybrid, higher-order, and Stokes dynamics models \citep{Chengetal2017,Robinsonetal2022}, or on lengthening their steps \citep{LofgrenAhlkronaHelanow2021}.

However, actual \emph{implicit} time-stepping \citep{LeVeque2007} should also be considered.  Here the velocity and geometry are updated simultaneously by solving coupled mass and momentum conservation equations.  While implicit time-stepping always requires the solution of systems of equations at each step, for ice sheets an implicit step must simultaneously compute the velocity and the \emph{domain on which the velocity is defined}, namely the 3D extent of the ice once the coupled solution has converged.  The problem is of free-boundary type, in map-plane (horizontal) directions \citep{SchoofHewitt2013}, and also in easily-resolved vertical directions.

An implicit strategy has been demonstrated at high-resolution in the simplest frozen-base, isothermal SIA case \citep{Bueler2016,JouvetGraeser2013}.  \cite{Bueler2016} also shows how the steady-state problem \citep{JouvetBueler2012} can often be solved, demonstrating unconditional stability.  (Observe that steady-state equations correspond to an implicit time step of infinite duration.)

The corresponding problem for the Stokes equations has not, to the author's knowledge, yet been attempted.  However, important early work applying a semi-implicit time step using Stokes dynamics \citep{WirbelJarosch2020} to solve a free boundary problem illuminates some of the techniques and difficulties needed to make such a strategy work for a membrane-stress-resolving balance.

This is the context in which the current note relates time-stepping and stress-balance choices to computational effort.  Our simplified performance analysis exposes the most important considerations and trade-offs.  While the author expects that implicit time-stepping ice sheet models will eventually be the fastest, and that advanced solver techniques like multigrid \citep{Briggsetal2000} will be necessary, these beliefs should be assessed quantitatively to the extent possible.

\section{Mass continuity equation}

Reader familiarity is assumed with the standard SIA and Stokes stress-balance equations \citep{GreveBlatter2009,SchoofHewitt2013}.  These (continuum) models are regarded here as end members of current-usage stress-balance approximations \citep{Robinsonetal2022}.  Familiarity with the mass continuity and surface kinematical equations \citep{GreveBlatter2009} is also assumed.  However, before analyzing performance, the form of the mass continuity equation must be examined.

For an incompressible ice sheet with thickness $H(t,\bx)$, vertically-averaged horizontal velocity $\bu(t,\bx)$, and climatic-basal mass balance $a(t,\bx)$, this equation says
\begin{equation}
\frac{\partial H}{\partial t} + \Divx \left(\bu H\right) = a, \label{eq:masscontinuity}
\end{equation}
where $\bx=(x,y)$ denotes horizontal coordinates.

Equation \eqref{eq:masscontinuity} suggests that ice sheets change geometry in an essentially advective manner, but this appearance is deceiving, or at least over-simplified, especially regarding the growth of numerical instabilities.  This is because ice flows dominantly downhill.  Indeed, ice sheet flow has no characteristic curves, as would \eqref{eq:masscontinuity} if it were a true advection, because the velocity $\bu$ actually depends on the gradient of thickness through the stress balance.

Thus, as can be addressed by linearized analysis \citep{Robinsonetal2022}, when thickness perturbations grow unstably under explicit time-stepping, i.e.~with too large a step, they do so by a mix of (discretized) advective and diffusive mechanisms.  Let $s(t,\bx)=H(t,\bx)+b(\bx)$ denote the surface elevation, for bed elevation $b(\bx)$.  Under a numerical thickness perturbation the velocity $\bu$ will often respond by increasing in a direction close to downhill ($-\gradx s$), a direction correlated to $-\gradx H$ over large areas of an ice sheet.  In membrane-stress-resolving models like Stokes this happens through the non-local solution of the stress balance, in which the gravitational source term is along $\gradx s$.  Equation \eqref{eq:masscontinuity} has velocity $\bu(H,\grad_x s)$ which is a non-local function of geometry, and a stress balance solution is required to evaluate it.  A numerical instability occurs when the ice thickness under/over-shoots its correct value because the numerical velocity response from evaluating this non-local function is too strong for the given time step.

\newcommand{\nn}{\text{n}}
A diffusive description of the mass continuity equation is valid in the small-aspect-ratio limit which generates the SIA \citep{SchoofHewitt2013}:
\begin{equation}
\frac{\partial H}{\partial t} = \Divx \left(d\, \gradx s \right) + a. \label{eq:siamasscontinuity}
\end{equation}
Here $d = C H^{\nn+2} |\gradx s|^{\nn-1}$ is the nonlinear diffusivity.\footnote{In detail, $C = 2 A (\rho g)^\nn/(\nn+2)$ in the isothermal case, where $A$ is the ice softness, $\rho$ is the ice density, $g$ is gravity, and $\nn\approx 3$ is the Glen exponent in the flow law \citep{GreveBlatter2009}.}  While equation \eqref{eq:siamasscontinuity} does not hold for Stokes or other membrane-stress-resolving dynamics, the same diffusivity $d$, an essentially geometric quantity, can be computed.  Generically, across stress balance choices, for grounded ice sheets one observes that large values of $d$ tend to indicate locations of unstable mode growth if explicit time-steps are chosen too large.

\begin{table*}[ht]
{\normalsize
\begin{tabular}{cll}
\emph{name} & \emph{meaning} & \emph{units} \\ \hline
$\alpha$    & one fixed-geometry Stokes velocity solution requires $O(n^{1+\alpha})${\large \strut} work\\
$\beta$     & one implicit SIA geometry-update (and velocity) solution requires $O(n^{1+\beta})$ work \\
$\gamma$    & one implicit, coupled Stokes geometry-update and velocity solution requires $O(n^{1+\gamma})$ work \\
$D$         & representative geometric (SIA) diffusivity of an ice sheet & $\text{km}^2 \text{a}^{-1}$ \\
$L$         & width of map-plane domain & km \\
$m$         & degrees of freedom: number of nodes in the horizontal mesh \\
$q$         & time steps per model year needed to resolve modeled climate interactions & $\text{a}^{-1}$ \\
$\Delta t$  & length of time step in model years & a \\
$U$         & representative horizontal ice velocity & $\text{km}\,\text{a}^{-1}$ \\
$\Delta x$  & representative width (diameter) of map-plane mesh cells & km
\end{tabular}
}
\caption{Parameters for performance analysis.  Note $\alpha$, $\beta$, $\gamma$, and $m$ are pure numbers.}
\label{tab:notation}
\end{table*}

\section{Performance analysis}

Table \ref{tab:notation} lists the parameters used in our performance analysis.  The primary parameters are $\Delta x$, a representative value for the horizontal mesh (grid) cell diameter, and $m$, the number of nodes (vertices) in the horizontal mesh.  \emph{High resolution} refers to the equivalent $\Delta x\to 0$ and $m\to\infty$ limits, and big-O notation is only used in this limit.

If the map-plane model domain is of width $L$ then these primary parameters are related by
\begin{equation}
\Delta x = O\left(\frac{L}{\sqrt{m}}\right) \quad \text{and} \quad m = O\left(\frac{L^2}{\Delta x^2}\right).  \label{eq:paramrelation}
\end{equation}
For a given domain width $L$ there are $O(\sqrt{m})$ mesh cells in each horizontal dimension.\footnote{In flow-line models $\Delta x = O(L m^{-1})$.  However, our analysis addresses spatially-3D models with map-plane horizontal meshes.}

A numerical ice sheet model uses $m$ ice thickness or surface elevation variables, one \emph{degree of freedom} per mesh node.  Storing these model state variables, plus the thermodynamical state, requires $O(m)$ memory if the mesh/grid has \emph{a priori} bounded resolution in the vertical direction.  The amount of fast memory needed by the simulation is also $O(m)$ if prior states are discarded or transferred to storage.  Such ice sheet models also have $O(m)$ velocity variables, but these are not state variables.  That is, a very-viscous stress balance computes velocity as a function of the true state variables.

Our assumption of fixed vertical resolution reflects common usage \citep[for example]{Aschwandenetal2019,BrinkerhoffJohnson2015,Hoffmanetal2018,Lengetal2012,
Winkelmannetal2011}, and it permits a rational comparison of asymptotics, but it is not the only possibility.  Some solvers use 3D refinement \citep{BrownSmithAhmadia2013,IsaacStadlerGhattas2015,Tuminaroetal2016}, with various distinctions between how horizontal and vertical meshing is handled.  This and many other details cannot be pursued here.

Ice sheet models resolve climate interactions, especially via surface mass balance, on time scales which are dominated by an annual cycle, and longer scales.  Let $q$ be the number of ice-dynamical time steps per year needed to capture this coupling.  Typical values $q=0.1 \,\text{a}^{-1}, 1 \,\text{a}^{-1}, 12 \,\text{a}^{-1}$ correspond to decadal, yearly, and monthly frequency, respectively.  Note that energy balance and degree-day schemes for computing surface mass balance generally have much shorter time scales, but here $q$ describes the frequency on which ice geometry is updated using ice velocity, i.e.~via solution of the mass continuity or surface kinematical equation.

Current-technology ice sheet models use explicit time-stepping which is only conditionally stable.  For the spatial resolutions used in present-day scientific applications, we observe that maintenance of explicit time-stepping stability requires steps substantially shorter than $1/q$ model years.

For an explicit SIA model the well-known stability restriction is $\Delta t < O(D^{-1} \Delta x^2)$ \citep{Bueleretal2005,HindmarshPayne1996} where $D$ is a representative diffusivity value, i.e.~of the $d$ in equation \eqref{eq:siamasscontinuity}.  For Stokes or other membrane-stress-resolving dynamics the stability of explicit time-stepping is largely unexplored in any precise sense, but an advective restriction $\Delta t < O(U^{-1} \Delta x)$, for some representative horizontal velocity scale $U$, represents the \emph{optimistic} paradigm.  The corresponding \emph{pessimistic} paradigm requires $\Delta t < O(D^{-1} \Delta x^2)$, using a representative diffusivity value $D$ computed in the SIA manner.

Explicit time-stepping with hybrid and higher-order schemes is somewhat better-studied than for Stokes dynamics, especially over horizontal resolutions relevant to whole ice sheets.  Some hybrid schemes apply the pessimistic paradigm as an adaptive restriction \citep{Winkelmannetal2011}.  Other models apparently require the user to choose a fixed time step through trial and error in some circumstances \citep[for example]{Fischleretal2022,Robinsonetal2022}.  The optimistic paradigm is supported in theory for a certain higher-order DIVA scheme \citep[see equation (52)]{Robinsonetal2022}, but practical Greenland simulations in the same work actually suggest an intermediate power $\Delta t = O(\Delta x^{1.6})$ (Figure 3(a)).

Unconditionally-stable implicit schemes also have a maximum time step restriction, namely $\Delta t < O(q^{-1})$.  This restriction reflects the simulation purpose, not maintenance of stability.  For an implicit scheme the frequency of climate coupling determines the total simulation cost, according to the (large) solution cost, at each time step, of coupled mass and momentum equations.

For each explicit time step of a model using a membrane-stress-resolving balance the computational cost of a velocity (or velocity/pressure) solution of the stress balance equations is determined by solver design.  We suppose that one such solution requires $O(m^{1+\alpha})$ floating point operations (flops),  with the power $\alpha\ge 0$ depending on the solver implementation.  For example, a Stokes solver using direct linear algebra for each Newton step might yield $\alpha \approx 1$ if sparsity is exploited or $\alpha \approx 2$ if not \citep{Bueler2021}.

% see data/scaling.m for results in next paragraphs

By contrast, a multigrid method \citep{Trottenbergetal2001} can greatly reduce $\alpha$.  For example, Antarctic ice sheet results by \cite{IsaacStadlerGhattas2015}, for a Stokes solver implemented using algebraic multigrid, show that the total number of preconditioned Krylov iterations \citep{Bueler2021}, over the nonlinear solve, grows slowly under mesh refinement, suggesting perhaps $\alpha\approx 0.2$ \citep[Table 8.1]{IsaacStadlerGhattas2015}.  The \cite{Lengetal2012} algebraic-multigrid Stokes solver may have similar scaling, but reported results do not constrain $\alpha$.  For a higher-order stress balance on the Greenland ice sheet, \cite{Tuminaroetal2016} report total algebraic-multigrid-preconditioned iterations suggesting $\alpha \approx 0.05$,\footnote{See Table 7.5.  Somewhat worse performance for the Antarctic ice sheet is diagnosed as caused by the difficulties in discretizing a marine margin.} and a geometric multigrid method by \citep{BrownSmithAhmadia2013} suggests $\alpha$ is close to zero for simplified geometries.

Note that $\alpha=0$ describes an \emph{optimal} solver in the language of algorithmic scaling or solver complexity \citep{Bueler2021}.  At the easiest end of ice sheet modeling, the SIA velocity computation is a trivialization of the Stokes problem in which velocity is computed by a pointwise formula; an SIA velocity solution therefore requires optimal $O(m)$ flops.

When analyzing solver scaling in our simplified form, one must be aware that the constant in $O(m^{1+\alpha})$ can be very large, strongly depending on solver design.  Furthermore, many considerations are suppressed in any flops-based analysis of algorithmic scaling.  Actual run time is also determined by memory latency, memory bandwidth, and process/thread/GPU parallelism, among other factors.  For example, the \cite{BrownSmithAhmadia2013,Fischleretal2022,IsaacStadlerGhattas2015,Lengetal2012,Tuminaroetal2016} results, among others, all show good parallel scaling, something we do not address here.

Regardless of the stress balance, an explicit time-stepping scheme then applies the mass continuity equation to update the ice thickness using $O(m)$ work.  That is, once the velocity is computed for the previous time-step's geometry, an explicit scheme replaces old thickness values by new ones using a pointwise formula.  Any additional computation needed to remesh the updated geometry, a design-dependent cost, is omitted here.

\newcommand{\oo}[1]{\displaystyle O\left(#1\right)}
\setlength{\tabcolsep}{5pt}
\renewcommand{\arraystretch}{1.5}
\begin{table*}[ht]
{\normalsize
\begin{tabular}{llll}
\emph{time-stepping} & \emph{dynamics} & \emph{flops per model year} & \emph{[pessimistic stability]} \\ \hline
explicit & SIA    & $\oo{\frac{D\, L^2}{\Delta x^4}} = \oo{\frac{D\, m^2}{L^2}}${\Huge \strut} \\
explicit & Stokes & $\oo{\frac{U L^{2+2\alpha}}{\Delta x^{3+2\alpha}}} = \oo{\frac{U m^{1.5+\alpha}}{L}}${\Huge \strut}\phantom{x} & $\oo{\frac{D\, L^{2+2\alpha}}{\Delta x^{4+2\alpha}}} = \oo{\frac{D\,m^{2+\alpha}}{L^2}}$ \\
implicit & SIA    & $\oo{\frac{q\, L^{2+2\beta}}{\Delta x^{2+2\beta}}} = \oo{q\, m^{1+\beta}}${\Huge \strut} \\
implicit & Stokes & $\oo{\frac{q\, L^{2+2\gamma}}{\Delta x^{2+2\gamma}}} = \oo{q\, m^{1+\gamma}}${\Huge \strut}
\end{tabular}
}
\caption{Asymptotic estimates of algorithmic scaling, measured by floating point operations per model year, for map-plane (2D) time-stepping numerical ice sheet simulations, in the high resolution limit where $\Delta x\to 0$ and $m\to\infty$.  See Table \ref{tab:notation} for notation.}
\label{tab:performancemodel}
\end{table*}

Now, how many flops are needed to simulate one model year?  Suppose a numerical model takes time steps of $\Delta t$ model years, equivalently $\Delta t^{-1}$ steps per model year.  We may write stability restrictions as required numbers of steps per model year.  That is, $\Delta t^{-1}$ is bounded below by a function of the horizontal resolution $\Delta x$ or the degrees of freedom $m$.  Recalling scaling \eqref{eq:paramrelation} of $\Delta x$ with $m$, in the explicit SIA and pessimistic-Stokes cases we have
\begin{equation}
\frac{1}{\Delta t} > \oo{\frac{D}{\Delta x^2}} = \oo{\frac{D m}{L^2}}. \label{eq:explicitsiarequired}
\end{equation}
The explicit, optimistic-Stokes estimate becomes
\begin{equation}
\frac{1}{\Delta t} > \oo{\frac{U}{\Delta x}} = \oo{\frac{U m^{1/2}}{L}}. \label{eq:explicitoptstokesrequired}
\end{equation}
The number of time steps per model year is then multiplied by the per-step computational cost, namely $O(m^{1+\alpha})$, to give a work estimate for each model year in a simulation.  The results so far are shown in the ``explicit'' rows of Table \ref{tab:performancemodel}.

As already explained, unconditionally-stable implicit methods have a fixed time step $\Delta t = 1/q$, independent of $\Delta x$ and determined only by the need to resolve climatic interactions.  On the other hand, the per-step expense is much greater because nontrivial coupled equations, indeed a free-boundary problem, must be solved for the velocity and geometry-update simultaneously.  For SIA models we parameterize the flops of such coupled solutions as $O(m^{1+\beta})$ with $\beta \ge 0$.  A large constant is assumed to be present.

% regarding next paragraph, see data/bueler.perf and data/scaling.m

The only implemented, unconditionally-stable, fully-implicit geometric-update solvers use the SIA stress balance.  For simplified dome geometry the scheme in \cite{Bueler2016}, based on Newton steps solved via direct linear algebra and single-grid ice margin determination, shows $\beta=0.8$.  (An earlier non-Newton implementation by \cite{JouvetBueler2012} scales worse.)  Convergence is robust for time steps of years to centuries on kilometer-scale grids for the Greenland ice sheet, using realistic, thus irregular, bed topographies.  Practical steady-state ($\Delta t=\infty$) solutions at ice sheet scale are also demonstrated by \cite{JouvetBueler2012} and \cite{Bueler2016}.

The SIA portions of hybrid time-stepping schemes by \cite{JouvetGraeser2013} and \cite{BrinkerhoffJohnson2015} are also solved implicitly.  \cite{JouvetGraeser2013} use a multigrid method, but their published, simplified-geometry results do not constrain $\beta$.  Overall time-stepping in these hybrids is only semi-implicit because the membrane-stress-resolving portion of the velocity explicitly advects the thickness, so the SIA-portion solvers are apparently only tested for time steps satisfying an advective condition $\Delta t < O(U^{-1}\Delta x)$ for $U$ which scales according to the sliding.
% equation (20) in JouvetGraeser2013
% equation (A21) in BrinkerhoffJohnson2015

For implicit Stokes time-stepping, a prospective, coupled, and free-boundary velocity and geometry-update solve is assumed to be, in the absence of constraining research, $O(m^{1+\gamma})$ for some $\gamma \ge \alpha$ to be determined.  One might also suppose $\gamma\ge \beta$, but in any case there are no implemented cases to measure.  These comments complete Table \ref{tab:performancemodel}, in which all estimates involve a scheme-dependent constant, something which is especially large for the implicit schemes.

% regarding the nMCD method proposed in [Bueler unpublished], which is expensive but multigrid, it may turn out that $\gamma$ is small

\section{Discussion and Conclusion}

From Table \ref{tab:performancemodel} we first observe a known property of explicit time-stepping for 2D (map-plane) diffusion equations such as SIA equation \eqref{eq:siamasscontinuity}, namely that effort, here flops per model year, scales as $O(\Delta x^{-4})$.  Recall that this follows because $\Delta t < O(\Delta x^2)$, and because the expense of one geometry-update operation is $O(m) = O(\Delta x^{-2})$.  Spatial mesh refinement by a factor of two therefore imposes an impressive 16-times increase in effort.

The \cite{Bueler2016} implicit SIA solver is not enough better, however.  Although long time steps can be taken by this implicit solver, the $\beta=0.8$ scaling gives $O(\Delta x^{-3.6})$ effort because $2+2\beta=3.6$.  However, an improvement to $\beta < 0.5$, presumably by application of a multigrid method, would transform such an implicit solver into a tool with notably superior performance compared to an explicit, $O(\Delta x^{-4})$ SIA scheme.

Now bypassing the over-simplified SIA stress balance, we see from Table \ref{tab:performancemodel} how the algorithmic scaling of solvers completely dominates.  In particular, an $\alpha=1$ explicit Stokes model, e.g.~one using direct, sparsity-exploiting linear algebra on each Newton step system, will do work which scales at the horrific rate $O(\Delta x^{-6})$ under a pessimistic stability condition.  Optimistic stability yields still-bad $O(\Delta x^{-5})$.  These simple observations emphasize the key role of multigrid-based stress balance solvers.  That is, even retaining explicit time-stepping, nearly-optimal ($\alpha \approx 0$) scaling of computation effort in Stokes and higher-order solvers will be necessary for routine application on high-resolution meshes.

On the other hand, suppose resolution ($\Delta x$) is fixed.  The Table also shows why algorithmic scaling remains important in the large-domain $L\to\infty$ limit when applying Stokes or higher-order dynamics.  An $\alpha=1$ method which might suffice for a smaller $L=100$ km ice cap will struggle for a $L=1000$ km ice sheet because the effort scales as the fourth ($2+2\alpha=4$) power of $L$.  By contrast, the computational work of a nearly-optimal solver will be proportional to ice sheet area $L^2$.

The promise of nearly-optimal solvers is profoundly revealed when implicit geometry updates are considered.  Future methods which simultanously satisfy the mass and momentum equations at each time step, and which do work essentially proportional to the number of degrees of freedom, have the greatest promise.  A $\gamma \approx 0$ implicit Stokes method would be hugely more capable for many tasks.  Specifically, and more achievably, a $\gamma < 0.5$ implicit, essentially unconditionally-stable, Stokes time-stepping method, presumably based on multigrid solution of the free-boundary problem for the coupled mass and momentum equations, is an appropriate goal for coming decades of research on numerical ice sheet models.  The computation cost would scale at $O(\Delta x^{-3})$, better than explicit SIA models, and the method would only update ice geometry when determined by the scientific need to resolve climate coupling, while avoiding shallow approximations.  The same goal makes sense for all membrane-stress-resolving solvers.  No apparent technical progress has yet been made on such an implicit Stokes design, and so these aspirations are decidedly long-term.  However, the above discussion suggests why measured values for $\alpha,\beta,\gamma$, or equivalent algorithmic scaling measures, are important performance metrics to report when describing new ice sheet solvers.

\subsubsection{Note added 4 July 2022.}  Errors in the first posted version (\url{https://arxiv.org/abs/2206.14352}) are corrected here.

%         References
\bibliography{perfmod}

\begin{thebibliography}{28}
\providecommand{\natexlab}[1]{#1}
\expandafter\ifx\csname urlstyle\endcsname\relax
  \providecommand{\doi}[1]{doi: #1}\else
  \providecommand{\doi}{doi: \begingroup \urlstyle{rm}\Url}\fi

\bibitem[\protect\citename{Aschwanden and others, }2019]{Aschwandenetal2019}
Aschwanden A, Fahnestock MA, Truffer M, Brinkerhoff DJ, Hock R, Khroulev C,
  Mottram R and Khan SA (2019) Contribution of the {Greenland Ice Sheet} to sea
  level over the next millennium. \emph{Science Advances}, \textbf{5}(6)

\bibitem[\protect\citename{Briggs and others, }2000]{Briggsetal2000}
Briggs W, Henson VE and McCormick S (2000) \emph{A {M}ultigrid {T}utorial}.
  SIAM Press, Philadelphia, 2nd edition

\bibitem[\protect\citename{Brinkerhoff and Johnson,
  }2015]{BrinkerhoffJohnson2015}
Brinkerhoff DJ and Johnson JV (2015) Dynamics of thermally induced ice streams
  simulated with a higher-order flow model. \emph{J. Geophys. Res.: Earth
  Surface}, \textbf{120}(9), 1743--1770

\bibitem[\protect\citename{Brown and others, }2013]{BrownSmithAhmadia2013}
Brown J, Smith B and Ahmadia A (2013) Achieving textbook multigrid efficiency
  for hydrostatic ice sheet flow. \emph{SIAM J. Sci. Compit.}, \textbf{35}(2),
  359--375

\bibitem[\protect\citename{Bueler, }2016]{Bueler2016}
Bueler E (2016) Stable finite volume element schemes for the shallow ice
  approximation. \emph{J. Glaciol.}, \textbf{62}(232), 230--242

\bibitem[\protect\citename{Bueler, }2021]{Bueler2021}
Bueler E (2021) \emph{{PETSc} for {P}artial {D}ifferential {E}quations:
  {N}umerical {S}olutions in {C} and {P}ython}. SIAM Press, Philadelphia

\bibitem[\protect\citename{Bueler and others, }2005]{Bueleretal2005}
Bueler E, Lingle CS, Kallen-Brown JA, Covey DN and Bowman LN (2005) Exact
  solutions and verification of numerical models for isothermal ice sheets.
  \emph{J. Glaciol.}, \textbf{51}(173), 291--306

\bibitem[\protect\citename{Cheng and others, }2017]{Chengetal2017}
Cheng G, L{\"o}tstedt P and von Sydow L (2017) Accurate and stable time
  stepping in ice sheet modeling. \emph{J. Comput. Phys.}, \textbf{329}, 29--47

\bibitem[\protect\citename{Fischler and others, }2022]{Fischleretal2022}
Fischler Y, R\"uckamp M, Bischof C, Aizinger V, Morlighem M and Humbert A
  (2022) A scalability study of the {Ice-sheet and Sea-level System Model}
  ({ISSM}, version 4.18). \emph{Geoscientific Model Development},
  \textbf{15}(9), 3753--3771

\bibitem[\protect\citename{Goelzer and others, }2020]{Goelzeretal2020}
Goelzer H, Nowicki S and others (2020) The future sea-level contribution of the
  {G}reenland ice sheet: a multi-model ensemble study of {ISMIP6}. \emph{The
  Cryosphere}, \textbf{14}(9), 3071--3096

\bibitem[\protect\citename{Greve and Blatter, }2009]{GreveBlatter2009}
Greve R and Blatter H (2009) \emph{Dynamics of {I}ce {S}heets and {G}laciers}.
  Advances in Geophysical and Environmental Mechanics and Mathematics,
  Springer, Berlin, Germany

\bibitem[\protect\citename{Hindmarsh and Payne, }1996]{HindmarshPayne1996}
Hindmarsh RCA and Payne AJ (1996) Time-step limits for stable solutions of the
  ice-sheet equation. \emph{Ann. Glaciol.}, \textbf{23}, 74--85

\bibitem[\protect\citename{Hoffman and others, }2018]{Hoffmanetal2018}
Hoffman MJ, Perego M, Price SF, Lipscomb WH, Zhang T, Jacobsen D, Tezaur I,
  Salinger AG, Tuminaro R and Bertagna L (2018) {MPAS-Albany Land Ice (MALI)}:
  a variable-resolution ice sheet model for {E}arth system modeling using
  {V}oronoi grids. \emph{Geosci. Model Dev.}, \textbf{11}(9), 3747--3780

\bibitem[\protect\citename{Isaac and others, }2015]{IsaacStadlerGhattas2015}
Isaac T, Stadler G and Ghattas O (2015) Solution of nonlinear {S}tokes
  equations discretized by high-order finite elements on nonconforming and
  anisotropic meshes, with application to ice sheet dynamics. \emph{SIAM J.
  Sci. Comput.}, \textbf{37}(6), B804--B833

\bibitem[\protect\citename{Jouvet and Bueler, }2012]{JouvetBueler2012}
Jouvet G and Bueler E (2012) Steady, shallow ice sheets as obstacle problems:
  well-posedness and finite element approximation. \emph{SIAM J. Appl. Math.},
  \textbf{72}(4), 1292--1314

\bibitem[\protect\citename{Jouvet and Gr{\"a}ser, }2013]{JouvetGraeser2013}
Jouvet G and Gr{\"a}ser C (2013) An adaptive {N}ewton multigrid method for a
  model of marine ice sheets. \emph{J. Comput. Physics}, \textbf{252}, 419--437

\bibitem[\protect\citename{Leng and others, }2012]{Lengetal2012}
Leng W, Ju L, Gunzburger M, Price S and Ringler T (2012) A parallel high-order
  accurate finite element nonlinear {S}tokes ice sheet model and benchmark
  experiments. \emph{J. Geophys. Res.: Earth Surface}, \textbf{117}(F1)

\bibitem[\protect\citename{LeVeque, }2007]{LeVeque2007}
LeVeque RJ (2007) \emph{{Finite Difference Methods for Ordinary and Partial
  Differential Equations: Steady-State and Time-Dependent Problems}}. SIAM
  Press, Philadelphia

\bibitem[\protect\citename{L{\"o}fgren and others,
  }2021]{LofgrenAhlkronaHelanow2021}
L{\"o}fgren A, Ahlkrona J and Helanow C (2021) Increasing stable time-step
  sizes of the free-surface problem arising in ice-sheet simulations, preprint
  arXiv:2106.16097

\bibitem[\protect\citename{Robinson and others, }2022]{Robinsonetal2022}
Robinson A, Goldberg D and Lipscomb WH (2022) A comparison of the stability and
  performance of depth-integrated ice-dynamics solvers. \emph{The Cryosphere},
  \textbf{16}(2), 689--709

\bibitem[\protect\citename{Schoof and Hewitt, }2013]{SchoofHewitt2013}
Schoof C and Hewitt IJ (2013) Ice-sheet dynamics. \emph{Annu. Rev. Fluid
  Mech.}, \textbf{45}, 217--239

\bibitem[\protect\citename{Seguinot and Delaney, }2021]{SeguinotDelaney2021}
Seguinot J and Delaney I (2021) Last-glacial-cycle glacier erosion potential in
  the {A}lps. \emph{Earth Surface Dynamics}, \textbf{9}(4), 923--935

\bibitem[\protect\citename{Seroussi and others, }2020]{Seroussietal2020}
Seroussi H, Nowicki S and others (2020) {ISMIP6 Antarctica}: a multi-model
  ensemble of the {A}ntarctic ice sheet evolution over the 21st century.
  \emph{The Cryosphere}, \textbf{14}(9), 3033--3070

\bibitem[\protect\citename{Trottenberg and others, }2001]{Trottenbergetal2001}
Trottenberg U, Oosterlee CW and Schuller A (2001) \emph{Multigrid}. Elsevier,
  Oxford, UK

\bibitem[\protect\citename{Tuminaro and others, }2016]{Tuminaroetal2016}
Tuminaro R, Perego M, Tezaur I, Salinger A and Price S (2016) A matrix
  dependent/algebraic multigrid approach for extruded meshes with applications
  to ice sheet modeling. \emph{SIAM J. Sci. Computing}, \textbf{38}(5),
  C504--C532

\bibitem[\protect\citename{Weber and others, }2021]{Weberetal2021}
Weber ME, Golledge NR, Fogwill CJ, Turney CSM and Thomas ZA (2021)
  Decadal-scale onset and termination of {A}ntarctic ice-mass loss during the
  last deglaciation. \emph{Nature Comm.}, \textbf{12}(1), 1--13

\bibitem[\protect\citename{Winkelmann and others, }2011]{Winkelmannetal2011}
Winkelmann R, Martin MA, Haseloff M, Albrecht T, Bueler E, Khroulev C and
  Levermann A (2011) {The Potsdam Parallel Ice Sheet Model (PISM-PIK)--Part 1:
  Model description}. \emph{The Cryosphere}, \textbf{5}(3), 715--726

\bibitem[\protect\citename{Wirbel and Jarosch, }2020]{WirbelJarosch2020}
Wirbel A and Jarosch AH (2020) Inequality-constrained free-surface evolution in
  a full {S}tokes ice flow model (\textit{evolve\_glacier v1.1}).
  \emph{Geoscientific Model Development}, \textbf{13}(12), 6425--6445

\end{thebibliography}
\bibliographystyle{igs}

\end{document}